\documentclass[prb,twocolumn,unsortedaddress,showpacs,aps]{revtex4}

\usepackage{graphicx}
\usepackage{bm}
\usepackage{amsmath}

\begin{document}

\newcommand{\Ce}{CeCoIn$_5$}
\newcommand{\Tc}{T_{\text{c}}}
\newcommand{\Bcii}{B_{\text{c2}}}
\newcommand{\wav}{\lambda_{\text{n}}}
\newcommand{\fq}{\phi_{\text{0}}}

\title{Hexagonal and Square Flux Line Lattices in \Ce}

\author{Morten~Ring~Eskildsen}
\email{morten.eskildsen@physics.unige.ch}
\affiliation{DPMC, University of Geneva, 24 Quai E.-Ansermet,
             CH-1211 Gen\`{e}ve 4, Switzerland}

\author{Charles~D.~Dewhurst}
\affiliation{Institut Laue-Langevin, 6 Rue Jules Horowitz,
             F-38042 Grenoble, France}

\author{Bart~W.~Hoogenboom}
\altaffiliation[Present address: ]
               {M. E. M\"{u}ller Institute for Structural Biology, Biozentrum,
                University of Basel, Klingelbergstrasse 70,
                CH-4056 Basel, Switzerland}
\affiliation{DPMC, University of Geneva, 24 Quai E.-Ansermet,
             CH-1211 Gen\`{e}ve 4, Switzerland}

\author{Cedomir~Petrovic}
\altaffiliation[Present address: ]
               {Department of Physics, Brookhaven National Laboratory,
                Upton, New York 11973}

\author{Paul~C.~Canfield}
\affiliation{Ames Laboratory and Department of Physics and Astronomy,
             Iowa State University, Ames, Iowa 50011}

\date{\today}

\begin{abstract}
Using small-angle neutron scattering, we have imaged the magnetic flux line
lattice (FLL) in the $d$-wave heavy-fermion superconductor \Ce. At low fields
we find a hexagonal FLL. Around $0.6$ T this undergoes what is very likely a
first-order transition to square symmetry, with the nearest neighbors oriented
along the gap node directions. This orientation of the square FLL is consistent
with theoretical predictions based on the $d$-wave order parameter symmetry.
\end{abstract}

\pacs{74.60.Ec, 74.70.Tx}

\maketitle

Recently, a whole new family of heavy fermion superconductors has been
discovered. It includes pressure-induced superconductivity in CeRhIn$_5$
\cite{hegger00}, ambient pressure superconductivity in CeIrIn$_5$
\cite{petrovic01a}, and superconductivity with the highest known $\Tc = 2.3$ K
for any heavy fermion at ambient pressure in \Ce \ \cite{petrovic01b}. The
Ce$M$In$_5$ ($M$ = Rh, Ir, Co) family exhibits several similarities to other
correlated electron superconductors such as high-$\Tc$ cuprates and crystalline
organic metals: their crystal structure consists of alternating units of
CeIn$_3$ and $M$In$_2$ stacked sequentially along the $c$ axis
\cite{grin79,haga01}, the superconducting state borders on a magnetically
ordered phase giving rise to competition \cite{hegger00} or coexistence of
magnetism and superconductivity \cite{pagliuso01,zapf01}. Finally there is
evidence from thermal conductivity measurements \cite{izawa01} and NMR
\cite{kohori01} indicating that \Ce \ is a $d$-wave superconductor, with line
nodes along the [110] and [1\={1}0] directions $(d_{x^2 - y^2})$.
Theoretically, $d$-wave pairing is expected to stabilize a square flux line
lattice (FLL) \cite{berlinsky95,xu96,shiraishi99,ichioka99}, which was indeed
recently reported in the high-$\Tc$ superconductor
La$_{1.83}$Sr$_{0.17}$CuO$_{4+\delta}$ (LSCO) \cite{gilardi02}. However, in the
latter case with an orientation rotated 45$^{\circ}$ with respect to
theoretical predictions \cite{berlinsky95,xu96,shiraishi99,ichioka99}. In
addition, it is worth pointing out that studies of the FLL symmetry in LSCO
as well as in YBa$_2$Cu$_3$O$_{7-\delta}$ (YBCO) are susceptible to potential
complications in interpretation due to the orthorhombic crystal structure,
which leads to formaion of twin planes which can pin the FLL \cite{keimer93}.
On the other hand, the crystal structure of \Ce \ is tetragonal which excludes
twinning, and this material may therefore turn out to be a better example of a
``typical'' $d$-wave superconductor.

Here we report FLL imaging in \Ce, obtained by small-angle neutron scattering
(SANS). The FLL undergoes what appears to be a first-order, field driven
transition from a hexagonal to a square FLL is observed. A square FLL has not
previously been observed in a heavy fermion superconductor, and furthermore
this is the first example of a square FLL in a $d$-wave superconductor oriented
with the nearest neighbor directions parallel to the nodal directions of the
gap.

The SANS experiment was carried out at the D11 small-angle neutron scattering
diffractometer at the Institut Laue-Langevin, Grenoble, France. Single crystals
of \Ce \ were grown from an excess indium flux \cite{petrovic01b}, and had a
$\Tc = 2.3$ K and $\Bcii(0) = 5.0$ T parallel to the $c$-axis. The sample
was composed of four single crystals with thicknesses $t = 0.16 - 0.2$ mm
mounted side by side, each of which was individually aligned. The rather thin
samples were necessary, due to the strong absorbtion of low-energy neutrons by
In. The total mass of the sample was 36 mg. Incident neutrons with wavelength
$\wav = 0.6$ nm and a wavelength spread $\Delta\wav/\wav = 10$\% were used, and
the FLL diffraction pattern was collected by a $64 \times 64$ (1 cm$^2$)
position sensitive detector. For all measurements, the sample was field cooled
to 50 mK in a dilution refrigerator insert, placed in a cryostat with a
superconducting magnet. Magnetic fields in the range $0.3$ to 2 T were applied
parallel to the crystalline $c$-axis, and background subtraction was performed
using measurements following a zero-field cooling.

\begin{figure*}
\includegraphics{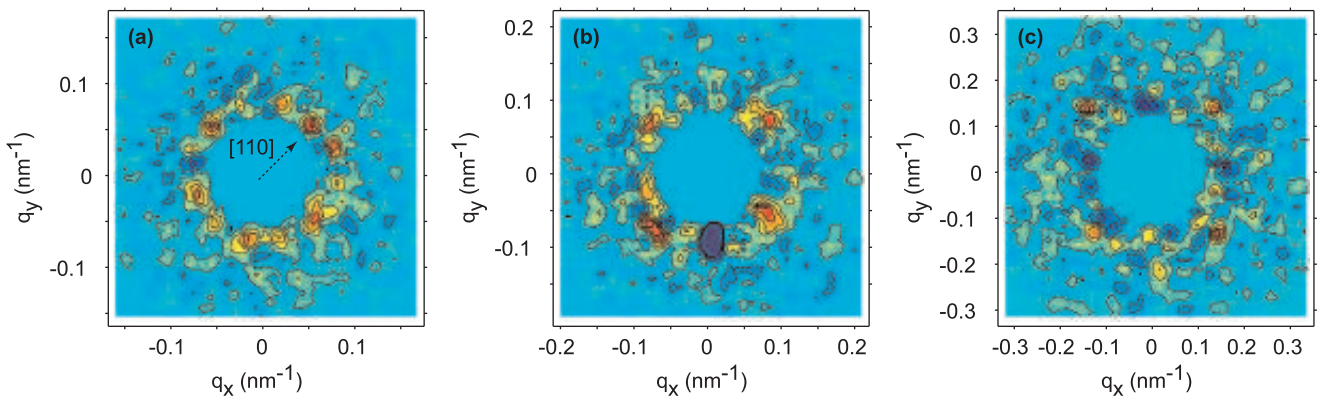}
\caption{(color)
         FLL diffraction patterns for \Ce \ obtained at 50 mK and applied
         fields of $0.3$ T (a), $0.6$ T (b) and 2 T (c), after subtraction of
         background measurement. In addition the data are smoothed by a
         $3 \times 3$ boxcar average, and the center of the image is masked
         off. The arrow in panel (a) shows the orientation of the crystalline
         axes.
         \label{FLL}}
\end{figure*}

\begin{figure}[b]
\includegraphics{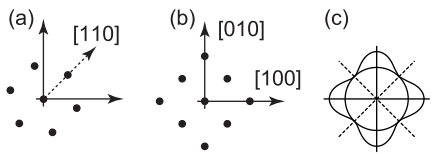}
\caption{Symmetry and orientation of the hexagonal (a) and square (b) FLL.
         The orientation of the nodes in the superconducting gap are shown in
         (c).
         \label{schem}}
\end{figure}

In Fig. \ref{FLL} we show FLL diffraction patterns for applied fields of $0.3$,
$0.6$ and 2 T.
The images were constructed by summing a number of measurements at different
angular positions, in order to satisfy the Bragg condition for the different
peaks. A clear evolution of the FLL symmetry and orientation is evident.
At the lowest field, twelve peaks are observed evenly distributed on a circle
in reciprocal space as shown in Fig. \ref{FLL}(a). This corresponds to two
hexagonal domains oriented along the [110] or the [\={1}10] directions. For a
hexagonal FLL oriented with respect to an underlying square crystal symmetry,
the existence of two degenerate domain orientations having equal population is
expected. As the field is increased to $0.6$ T, a primarily square FLL is found
[Fig. \ref{FLL}(b)], with the majority of the scattered intensity concentrated
in four irregularly shaped peaks. Again the FLL is oriented with the nearest
neighbor direction along [110] and [\={1}10].
The square FLL remains stable up to the highest measured field of 2 T [Fig.
\ref{FLL}(c)], where the diffraction pattern now shows four distinct Bragg
peaks.
The symmetry and orientation of the FLL is shown schematically in real space
in Fig. \ref{schem}, and compared to the symmetry of the superconducting gap.
While preserving the nearest neighbor direction along [110], the transition
from hexagonal to square symmetry can in principle be continous. However, at
intermediate fields this would result in diffraction patterns with
contributions from 4 sheared hexagonal lattice orientations, analogous to
what was previously observed in YBCO \cite{keimer93}. Such a distortion of the
hexagonal FLL was not observed, and the transition from square to hexagonal
symmetry is therefore most likely discontinous, i.e. of first order. On the
other hand, we cannot exclude that the transition to square symmetry is
preceeded by a weak rhombic distortion of the hexagonal FLL, similar to what
was observed in the borocarbides \cite{eskildsen97a}. However, such a
distortion would not alter the order of the transition from being first order.
Finally, a co-existence of domains having respectively square and hexagonal
symmetry is expected and usually observed in a narrow field range around a
first-order transition \cite{levett02}. We expect this to be the reason for the
slightly disordered square diffraction pattern seen in Fig. \ref{FLL}(b), and
hence take the corresponding applied field of $0.6$ T to be at or close to the
transition field. The most likely first-order FLL symmetry transition, and the
orientation of the square FLL along the nodes of the superconducting gap, are
the main results of this report, and will be addressed in further detail below.

\begin{figure}[b]
\includegraphics*{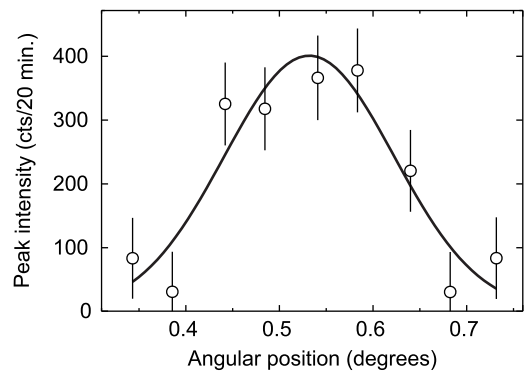}
\caption{Rocking curve for the \Ce \ (1,0) FLL reflection at 50 mK and 2 T.
         The curve is a fit to a Gaussian.
         \label{RC}}
\end{figure}

SANS FLL imaging in \Ce \ is complicated by two factors, illustrated by the
diffraction patterns in Fig. \ref{FLL} and the 2 T rocking curve shown in Fig.
\ref{RC}. The most limiting factor is the long superconducting penetration
depth in this material, which results in a very small field modulation and
hence low scattered intensity. This inevitably leads to imperfect background
subtractions, which is seen as negative scattered intensity in the diffraction
patterns in Fig. \ref{FLL} (dark blue regions).
Fig. \ref{RC} shows the intensity of a FLL Bragg reflection, as the cryostat is
gradually tilted and rotated in such a way that the scattering vector cuts the
Ewald sphere at a right angle.
The integrated reflectivity is given by \cite{eskildsen97b}
$$
R = \frac{2 \pi \gamma^2 \wav ^2 t}{16 \fq^2 q} \left| h(q) \right| ^2,
$$
where $\gamma = 1.91$ is the neutron gyromagnetic ratio, $t$ is the sample
thickness, $\fq = h/2e = 2067$ T\,nm$^2$ is the flux quantum, and
$q = 2 \pi \surd (B/\fq) = 0.1955 \mbox{ nm}^{-1}$ is the calculated scattering
vector for a square FLL and a field of 2 T. This expectation and the measured
value of $q = 0.19$ nm$^{-1}$ agree within 3\%. The flux line form factor
$h(q)$ for a square lattice is given by \cite{eskildsen97b}
$$
h(q) = \frac{\fq}{(2 \pi \lambda)^2} e^{-\pi B/\Bcii},
$$
where the exponential factor represents the so-called core correction. Fitting
the rocking curve to a Gaussian and using the area under the curve as the
integrated reflectivity, together with the upper critical field
$\Bcii(0) = 5.0$ T,\cite{petrovic01b} we obtain $\lambda = 247 \pm 10$ nm.
This falls inside the range of values $\lambda_0 = 190 - 281$ nm reported in
the litterature \cite{ormeno02,ozcan02,chia02}.
Calculating the coherence length from the upper critical field,
$\xi = \surd(\fq/2 \pi \Bcii)  = 8.1$ nm, we estimate the GL parameter
$\kappa = \lambda/\xi \approx 30$, making \Ce \ a strongly type-II
superconductor.

The second complicating factor is the narrow rocking curve, necessitating a
very precise alignment in order to obtain scattering. The fit to the data in
Fig. \ref{RC} yields a width of $0.21^{\circ}$ FWHM comparable to the
experimental resolution estimated to be $0.15^{\circ}$ FWHM. In principle the
rocking curve width can be used to determine the longitudinal correlation
length or straightness of the flux lines, but with the width being close to the
experimental resolution this should rather be taken as a lower bound. We find
$\Delta q_{\text{L}} \leq 0.21^{\circ} (\pi/180^{\circ}) q
                     =    7 \times 10^{-4} \mbox{ nm}^{-1}$,
and hence $\xi_{\text{L}} = 2/\Delta q_{\text{L}} \ge 3$ $\mu$m. This is a
large value corresponding to $\sim 100$ flux line spacings, and indicates very
weak pinning in this material.

We now return to the discussion of the symmetry and orientation of the FLL.
In an ideal isotropic type-II superconductor this will be hexagonal
\cite{abrikosov57,kleiner64}. However, if one evaluates the free energy
difference between the hexagonal and square symmetry, this is found to be only
about 2\% \cite{kleiner64}. A relatively weak anisotropy is therefore capable
of changing this delicate balance, leading to a distorted hexagonal or a square
FLL. A number of theoretical studies have addressed the effect of $d$-wave
pairing on the structure and orientation of the FLL. As the field is increased
or temperature decreased, they consistently find that a square FLL is
stabilized, oriented with the nearest neighbor direction {\em along} the
direction of the gap nodes \cite{berlinsky95,xu96,shiraishi99,ichioka99}.
Determining the orientation of the hexagonal FLL is more difficult, since the
energy difference between the two configurations aligned 45$^{\circ}$ apart is
very small \cite{xu96,ichioka99}. Ichioka {\em et al.} \cite{ichioka99}
conclude that both the square and the hexagonal FLL are oriented with the
nearest neighbors along the node direction, with a first-order transition
separating the two symmetries. The transition field is predicted to be
$0.15 \times \Bcii$, which in the case of \Ce \ corresponds to $0.75$ T at
$T = 0$ K. This is in agreement with our results concerning the orientation of
the FLL as well as the nature of transition. Furthermore, their prediction of
the transition field is in fair agreement with our estimate of $\sim 0.6$ T.

In principle there is another mechanism that could be responsible for the FLL
symmetry transition: A four-fold Fermi surface anisotropy combined with
nonlocal electrodynamics due to the finite coherence length. Theoretically,
this was studied extensively by Kogan {\em et al.}, who used nonlocal
corrections to the London model to calculate the FLL free energy and thereby
determine the stable configuration as a function of the flux line density
\cite{kogan96,kogan97a,kogan97b}. This is the driving force behind the
transition between a low-field (distorted) hexagonal and a high-field square
FLL seen in the rare-earth nickelborocarbides \cite{eskildsen97a,eskildsen97b}
as well as in V$_3$Si \cite{yethiraj99}.
In the case of \Ce \ such an analysis has
not yet been carried out. Band structure calculations have been performed on
the isostructural compound CeIrIn$_5$ \cite{haga01}, and (partially) confirmed
on \Ce \ by measurements of de Haas - van Alphen oscillations \cite{settai01}.
The calculations show a Fermi surface with at least one sheet having a
four-fold anisotropy \cite{haga01}. However, this warps between two
orientations $45^{\circ}$ apart, and at present it is not clear what
implication this has on the FLL.

To summarize, we have studied the symmetry and orientation of the magnetic flux
line lattice in the $d$-wave superconductor \Ce \ using small-angle neutron
scattering. At low fields a hexagonal FLL was found, which undergoes which is
most likely a first-order transition to square symmetry around $0.6$ T. Though
the possibility of a Fermi surface anisotropy combined with nonlocal effects
can not be ruled out as the determining factor, our measurements agree well
with the predictions for a pairing-symmetry driven transition. In particular,
the nature of the transition, the field at which it occurs, and above all the
orientation of the square FLL with the nearest neighbors alinged parallel to
the node directions, are all consistent with being driven by the $d$-wave
symmetry of the order parameter.

We thank R. Cubitt for assistance with the SANS experiment, and J.-L. Ragazzoni
for his persistance in getting the dilution fridge running. Fruitfull
discussions with A. Knigavko, V. G. Kogan, K. Machida, P. Miranovi\'{c} and
J. Mesot are gratefully acknowledged.


\end{document}